\documentclass[twocolumn]{revtex4}
\usepackage{graphicx}

\newcommand{\be}{\begin{equation}}
\newcommand{\ee}{\end{equation}}
\newcommand{\ba}{\begin{eqnarray}}
\newcommand{\ea}{\end{eqnarray}}
\newcommand{\go}{\agt}
\newcommand{\lo}{\alt}
\def\bxi{{\mbox{\boldmath $\xi$}}}

\def\br{{\bf r}}
\def\bC{{\bf C}}

\def\bOmega{{\bf \Omega}}
\def\Oms{\Omega_s}
\def\Omo{\Omega_{\rm orb}}

\def\omi{\omega_\alpha}

\def\lr{l}

\begin{document}
\title{Resonant Tidal Excitations of Inertial Modes in Coalescing
Neutron Star Binaries}
\author{Dong Lai}
\email{dong@astro.cornell.edu}
\affiliation{Center for Radiophysics and Space Research,
Department of Astronomy, Cornell University, Ithaca, NY 14853}
\author{Yanqin Wu}
\affiliation{Department of Astronomy and Astrophysics, 
University of Toronto, 60 St. George Street, Toronto, ON M5S 3H8, Canada}

\begin{abstract}
We study the effect of resonant tidal excitation of inertial modes
in neutron stars during binary inspiral. For spin frequencies
less than $100$~Hz, the phase shift in the gravitational waveform
associated with the resonance is small and does not affect the
matched filtering scheme for gravitational wave detection.
For higher spin frequencies, the phase shift can become significant. 
Most of the resonances take place at orbital frequencies
comparable to the spin frequency, and thus significant phase shift
may occur only in the high-frequency band (hundreds of Hertz) 
of gravitational wave. The exception is a single odd-paity $m=1$ mode,
which can be resonantly excited for misaligned spin-orbit inclinations, 
and may occur in the low-frequency band (tens of Hertz) of gravitational 
wave and induce significant ($\gg 1$~radian) phase shift.
\end{abstract}

\maketitle

\section{Introduction}

Coalescing neutron star-neutron star (NS-NS) and neutron star-black
hole (NS-BH) binaries are the most promising sources of gravitational
waves (GWs) for ground-based detectors such as LIGO and VIRGO\cite{cutler02}.
The last few minutes of the binary inspiral produce GWs with
frequencies sweeping upward through the 10--1000~Hz range (LIGO's
sensitivity band)\cite{cutler93}.
Due to the expected low signal-to-noise ratios, 
accurate gravitational waveforms are required to serve as theoretical
templates that can be used to detect the GW signal from the noise
and to extract binary parameters from the waveform.

In the early stage of the inspiral, with the GW frequencies between
$10$~Hz to a few hundred Hz, it is usually thought that the NS can be
treated as a point mass, and tidal effects are completely
negligible. This is indeed the case for the ``quasi-equilibrium''
tides as the tidal interaction potential scales as $a^{-6}$ (where $a$
is the orbital separation)\cite{Laietal94}.
The situation is more complicated for the {\it resonant tides}: As two
compact objects inspiral, the orbit can momentarily come into resonance
with the normal oscillation modes of the NS.  By drawing energy from
the orbital motion and resonantly exciting the modes, the rate of
inspiral is modified, giving rise to a phase shift in the
gravitational waveform.  This problem was first studied 
\cite{Reisenegger94,Lai94}
in the case of non-rotating NSs where the
only modes that can be resonantly excited are g-modes (with typical
mode frequencies $\lo 100$~Hz). It was found that the effect is
small for typical NS parameters (mass $M\simeq 1.4M_\odot$ and
radius $R\simeq 10$~km) because the coupling between the g-mode and
the tidal potential is weak. Ho \& Lai (1999)\cite{holai99} 
studied the effect of NS rotation, and found that the g-mode
resonance can be strongly enhanced even by a modest rotation (e.g.,
the phase shift in the waveform $\Delta\Phi$ reaches up to $0.1$~radian 
for a spin frequency $\nu_s\lo 100$~Hz). 
They also found that for a rapidly rotating NS ($\nu_s\go
500$~Hz), f-mode resonance becomes possible (since the inertial-frame
f-mode frequency can be significantly reduced by rotation) and produces
a large phase shift. In addition, NS rotation gives rise to
r-mode resonance whose effect is appreciable only for very rapid
(near breakup) rotations. Recently, Flanagan \& Racine (2006)\cite{Flanagan06} 
studied the gravitomagnetic
resonant excitation of r-modes and and found that the post-Newtonian effect
is more important than the Newtonian tidal effect (and the phase shift
reaches $\sim 0.1$~radian for $\nu_s\sim 100$~Hz).
Taken together, these previous studies suggest that for
astrophysically most likely NS parameters ($M\simeq 1.4M_\odot$,
$R\simeq 10$~km, $\nu_s\lo 100$~Hz), tidal resonances have a
small effect on the gravitational waveform during binary inspiral.

A rotating NS also supports a large number of Coriolis-force driven
modes named inertial modes (also called rotational hybrid modes
or generalized r-modes; see, e.g., Refs.~\cite{Papaloizou81,Lockitch99,
Schenk02,Wu05a}), of which r-mode is a member.
The modes have frequencies of order the spin frequency, and as we show
below, they couple more strongly to the (Newtonian) tidal potential 
than the r-mode. It is therefore important to investigate 
the effect of inertial mode resonances on the GW phase evolution 
during binary inspiral --- this is the goal of this paper.

\section{Basic Equations for Tidal Resonance During Binary Inspiral}

We consider a NS of mass $M$, radius $R$ and spin $\bOmega_s$
in orbit with a companion $M'$ (another NS or a black hole).
The orbital radius $a$ decreases in time due to GW emission, 
$\Omega_{\rm orb}$ is the orbital angular frequency.
We allow for a general spin-orbit inclination angle $\Theta$
(the angle between $\bOmega_s$ and the orbital angular momentum
${\bf L}$). In the spherical coordinate system centered on $M$ 
with the $Z$-axis along ${\bf L}$, the gravitational potential produced 
by $M'$ can be expanded in terms of spherical harmonics: 
\be
U(\br,t)
 =-GM'\sum_{l{m'}}{W_{lm'}r^l\over a^{l+1}} e^{-im'\Phi(t)}
Y_{lm'}(\theta_L,\phi_L), \label{eq:potential}
\ee
where $\Phi(t)=\int^t dt\,\Omo$ is the orbital phase and \cite{Press77}
\ba
W_{lm'}& = & (-)^{(l+m')/2}\left[{4\pi\over 2l+1}(l+m')!(l-m')!
\right]^{1/2}\nonumber \\
&& \times \left[2^l\left({l+m'\over 2}\right)!\left({l-m'\over 2}
\right)!\right]^{-1},
\label{eq:wlm}
\end{eqnarray}
[Here the symbol $(-)^p$ is zero if $p$ is not an integer.]
Only the $l\ge 2$ terms are relevant, and the dominant ($l=2$) tidal
potential has $W_{2\pm 2} = (3\pi/10)^{1/2}$. Since 
it is most convenient to describe oscillation modes relative to 
the spin axis, we need to express the tidal potential in terms of 
$Y_{lm}(\theta,\phi)$, the spherical harmonic function defined in 
the corotating frame of the NS with the $z$-axis along $\bOmega_s$. 
This is achieved by the relation
\be Y_{lm'}(\theta_L,\phi_L)=\sum_{m}{\cal D}^{(\lr)}_{mm'}
(\Theta)Y_{lm}(\theta,\phi_s), \ee where the ${\cal D}^{(l)}_{mm'}$ is
the Wigner ${\cal D}$-function (e.g., Wybourne 1974), and
$\phi_s=\phi+\Oms t$.

The linear perturbation of the tidal potential on the NS is specified
by the Lagrangian displacement, $\bxi(\br,t)$, of a fluid element
from its unperturbed position.  In the rotating frame of the NS, the
equation of motion takes the form
\be
\frac{\partial^2 \bxi}{\partial t^2}+2\bOmega_s\times
\frac{\partial\bxi}{\partial t}+{\bC}\cdot\bxi=-\nabla U, 
\label{eq:eqnmotion2}
\ee
where $\bC$ is a self-adjoint operator (a function of the pressure and
gravity perturbations) acting on $\bxi$ (see, e.g., 
Ref.\cite{Friedman78a}).
A free mode of frequency $\omega_\alpha$ with $\bxi_\alpha(\br,t)
=\bxi_\alpha(\br)\,e^{-i\omega_\alpha t}\propto
e^{im\phi-i\omega_\alpha t}$ satisfies
\be
-\omi^2\bxi_\alpha-2i\omi\bOmega_s\times\bxi_\alpha+\bC\cdot
\bxi_\alpha=0,
\ee
where $\{\alpha\}$ denotes the mode index, which includes the 
azimuthal ``quantum'' number $m$. We carry out phase space
mode expansion\cite{Dyson79,Schenk02}
\be
\left[\begin{array}{c}
\bxi\\
{\partial\bxi/\partial t}
\end{array}\right]
=\sum_\alpha c_\alpha(t)
\left[\begin{array}{c}
\bxi_\alpha(\br)\\
-i\omi\bxi_\alpha(\br)
\end{array}\right].
\ee
Using the orthogonality relation\cite{Friedman78a}
$\langle\bxi_\alpha,2i\bOmega_s\times\bxi_{\alpha'}\rangle+
(\omega_\alpha+\omega_{\alpha'})\langle\bxi_\alpha,\bxi_{\alpha'}
\rangle=0$ (for $\alpha\neq \alpha'$), where 
$\langle A,B\rangle\equiv\int\!d^3x\,\rho\, (A^\ast\cdot B)$,
we find \cite{Schenk02}
\ba
{\dot c}_\alpha+i\omi c_\alpha &=& 
{i\over 2\varepsilon_\alpha}\langle\bxi_\alpha(\br),-\nabla U\rangle\nonumber\\
&=& \sum_{m'}f_{\alpha,m'}\,e^{im\Omega_s t-im'\Phi},
\label{eq:adot}\ea
with
\be
f_{\alpha,m'}={iGM'\over
2\varepsilon_\alpha}\sum_{l}{W_{lm'}\over a^{l+1}}{\cal D}^{(l)}_{mm'}
Q_{\alpha,lm},
\ee
where
\ba
&&Q_{\alpha,lm}\equiv\bigl\langle\bxi_\alpha,\nabla (r^lY_{lm})
\bigr\rangle,
\label{eq:Qdefine}\\
&& \varepsilon_\alpha\equiv
\omi+\langle\bxi_\alpha,i\bOmega_s\times\bxi_{\alpha}\rangle,
\ea
and we have used the normalization $\langle\bxi_\alpha,\bxi_\alpha\rangle
=1$. Recall that in Eqs.~(\ref{eq:adot}-\ref{eq:Qdefine}), the index $\alpha$
includes $m$.

Now consider the excitation of a specific mode with inertial-frame 
frequency $\sigma_\alpha=\omi+m\Omega_s$ by the potential
component $\propto e^{-im'\Phi}$ --- we call this $(\alpha,m')$-resonance. 
In the following we shall adopt the convention $m>0$ and $m'>0$.
We first consider the case of $\sigma_\alpha>0$ (i.e., the mode is prograde 
with respect to the spin in the inertial frame, although 
in the rotating frame the mode can be either prograde or retrograde, 
corresponding to $\omega_\alpha>0$ and $\omega_\alpha<0$ respectively).
The resonance occurs when 
\be
\sigma_\alpha=m'\Omega_{\rm orb}.
\ee
Note that the mode azimuthal index $m=m'$ for aligned spin-orbit
($\Theta=0$), but in general the $m'$-th potential can excite a mode
with a different $m$. Integrating Eq.~(\ref{eq:adot}) across the 
resonance, we find that the post-resonance amplitude is given by
\be
c_\alpha e^{i\omega_\alpha t}=\int\!\!dt\, f_{\alpha,m'}\,
e^{i\sigma_\alpha t -im\Phi}
\simeq f_{\alpha,m'}\!\left({2\pi
\over m'\dot\Omo}\right)^{1/2},
\ee
where $a,\Omo$ should be evaluated at the resonance radius $a_\alpha$,
at which $\sigma_\alpha=m'[G(M+M')/a_\alpha^3]^{1/2}$.  The mode
energy in the inertial frame is $2\sigma_\alpha\varepsilon_\alpha
|c_\alpha|^2$, where the factor 2 accounts for the fact that for each
$m>0,~\sigma_\alpha>0$ resonant mode (recall that our convention is
$m>0$), there is also an identical $m<0,~\sigma_\alpha<0$ resonant
mode, and they should be counted as the same mode. Thus the energy
transfer to the mode during the resonance is given by
\ba
&&\Delta E_{\alpha,m'}={G{M'}^2\over R}{GM\over R^3}
\left({\pi\over m'\dot\Omega_{\rm orb}}\right)
{\sigma_\alpha\over\varepsilon_\alpha}\nonumber\\
&&\qquad \times \left[\sum_lW_{lm'}{\cal D}_{mm'}^{(l)}Q_{\alpha,lm}
\left({R\over a_\alpha}\right)^{l+1}\right]^2.
\label{eq:deltaE}\ea
where we have defined $Q_{\alpha,lm}$ in units such that $M=R=1$.
Except for differences in notation and sign convention,
Eq.~(\ref{eq:deltaE}) agrees with the expression derived in Ref.
\cite{holai99}, where mode decomposition was not carried out 
rigorously \cite{comment,lai97}.

The phase shift in the gravitational waveform due to the resonant
energy transfer is twice the orbital phase shift and is given by
\be 
\Delta\Phi_{\rm GW}=-2\Omo t_{\rm GW}{\Delta
E_{\alpha,m'}\over |E_{\rm orb}|},
\ee
where $E_{\rm orb}=-GMM'/(2a)$ is the orbital energy, 
and 
\be
t_{\rm GW}=|a/\dot a|={5c^5a^4\over G^3M^3q(1+q)}
\ee
is the orbital decay timescale ($q=M'/M$ is the mass ratio;
all quantities should be evaluated at the resonance radius $a_\alpha$). 
Keeping the
leading $l$-term in Eq.~(\ref{eq:deltaE}), we find
\ba
&&\Delta\Phi_{\rm GW}=-\frac{25\pi}{1536}\left(\frac{Rc^2}{GM}
\right)^5\frac{1}{q(1+q)^{(2\lr-1)/3}}\nonumber\\
&&\qquad \times\frac{1}{\hat \varepsilon_\alpha}
\left(\frac{\hat\sigma_\alpha}{m'}\right)^{(4 l-11)/3}\!\!\!
\left(W_{lm'}{\cal D}^{(l)}_{mm'}
Q_{\alpha,lm}\right)^2, \label{eq:orbchange}
\ea
where $\hat\sigma_\alpha=\sigma_\alpha (R^3/GM)^{1/2}$
and $\hat\varepsilon_\alpha=\varepsilon_\alpha (R^3/GM)^{1/2}$.

For modes with $\sigma_\alpha<0$ (i.e., retrograde with respect to spin
in the inertial frame), the GW phase shift at the resonance $-\sigma_\alpha
=m'\Omega_{\rm orb}$ can be similarly derived and is given by
\ba
&&\Delta\Phi_{\rm GW}=\frac{25\pi}{1536}\left(\frac{Rc^2}{GM}
\right)^5\frac{1}{q(1+q)^{(2\lr-1)/3}}\nonumber\\
&&~~\times\frac{1}{\hat \varepsilon_\alpha}
\left(\frac{|\hat\sigma_\alpha|}{m'}\right)^{(4 l-11)/3}\!\!\!\!
\left(W_{l,-m'}{\cal D}^{(l)}_{m,-m'}
Q_{\alpha,lm}\right)^2. \label{eq:orbchange2}
\ea

\section{Inertial Modes and Tidal Coupling Coefficients}

To fix notations, we first summarize the basic property of
inertial modes for incompressible stars\cite{Wu05a}.
To order ${\cal O}(\Omega_s)$, the mode displacement vector
$\bxi_\alpha\,e^{-i\omega t}\propto e^{im\phi-i\omega t}$
satisfies the equation
\be
\bxi_\alpha+iq({\bf e}_z\times\bxi_\alpha)=\nabla\psi,
\ee
where ${\bf e}_z$ is the unit vector along the z-axis ($\bOmega_s$),
$q=2\Omega_s/\omega$, and $\omega^2\psi=\delta P/\rho$ is the
(Eulerian) enthalpy perturbation.  For incompressible fluid,
$\nabla\cdot\bxi_\alpha=0$, we have 
\be 
\nabla^2\psi-q^2{\partial^2\psi\over\partial z^2}=0.
\label{eq:modeeqn}
\ee
This equation can be solved in an ellipsoidal coordinates $(x_1,x_2)$, 
giving $\psi\propto P_j^m(x_1)P_j^m(x_2)$.
For a given set of $(m,j)$,
there are $(j-m)$ eigenvalues $\omega_\alpha/\Omega_s$. The modes
with even $(j-m)$ have even parity with respect to the equator,
while those with odd $(j-m)$ have odd parity. In this notation,
``pure'' r-modes correspond to those with $j-m=1$ and have frequencies
$\omega_\alpha=-2\Omega_s/(1+m)$. Such characterization of inertial
modes can be generalized to stellar models which are compressible.

\subsection{Even-Parity Modes}

The $m=2$, even-parity ($j=4,6,\cdots$) inertial modes can be
excited by the $l=m'=2$ tidal potential. The 
frequencies of some modes are listed in Table I.
Consider the two $m=2,~j=4$ inertial modes. For uniform stellar models,
both have eigenfunctions
\be
\psi\propto \varpi^2\left[42\mu^2 z^2+7
(1-\mu^2)\varpi^2-6\right],
\ee
where $\mu=q^{-1}=\omega/(2\Omega_s)$, and $(\varpi,\phi,z)$ are 
cylindrical coordinates. A direct calculation shows that 
\ba
Q_{\alpha,22}&=&\int\!d^3\!x\,\rho\,\bxi_\alpha^\ast\cdot\nabla
(r^2Y_{22})\nonumber\\
&=&\int\!d^3\!x\,\rho\,{2\over 1+q}\left(\varpi{\partial\psi
\over\varpi}+2\psi\right)=0.
\ea
Thus, to order ${\cal O}(\Omega_s)$, the tidal coupling coefficient
vanishes.
%
The first non-zero contribution arises in the order ${\cal
O}(\Omega_s^2)$. Integration by part of equation (\ref{eq:Qdefine})
yields, 
\be 
Q_{\alpha,22}=\int\!d^3\!x\,\delta\rho_\alpha^\ast\,
r^2Y_{22}(\theta,\phi) + \oint r^2 Y_{22} \rho{\bxi}_{\alpha}^\ast
\cdot d{\hat S},
\label{eq:qalpha}
\ee
where $\delta\rho_\alpha$ is the Eulerian density perturbation and
$\delta\rho_\alpha=(\omega_\alpha^2 \rho^2/ \Gamma_1 P)\psi$ with
$\Gamma_1$ being the adiabatic index. Formally, $\Gamma_1 =
\infty$ in an incompressible model. The first term on the right-hand-side 
of Eq.~(\ref{eq:qalpha}) scales as 
$\Omega_s^2$. The Lagrangian pressure perturbation vanishes at the
surface, or $\delta P+\bxi\cdot\nabla P=0$. So the surface radial
displacement $\xi_{\alpha,r}$ satisfies
$\xi_{\alpha}=(\omega_\alpha^2/g)\psi(r=R) \propto
\Omega_s^2$ (where $g=GM/R^2$). As a result, the surface integral on the 
right-hand-side also comes in at the order of ${\cal O}(\Omega_s^2)$.

To accurately calculate tidal coupling to order ${\cal
O}(\Omega_s^2)$, it will be necessary to compute the mode
eigenfunction to order ${\cal O}(\Omega_s^2)$, including rotational
distortion to the hydrostatic structure and the centrifugal force in
the perturbation equation. This is beyond the scope of our paper. To
estimate the tidal coupling coefficient, we consider a compressible,
but uniform stellar model. The Eulerian density perturbation
$\delta\rho_\alpha$ is related to $\psi$ by
$\delta\rho_\alpha=(\omega_\alpha^2 \rho^2/
\Gamma_1 P)\psi$, and we fix $\Gamma_1$ to a finite value.
This $\delta\rho_\alpha$ is then used in Eq.~(\ref{eq:qalpha}) to
obtain $Q_{\alpha,22}$. The results are given in Table I.

\begin{table}
\caption{Frequencies and Tidal Coupling Coefficients for $m=2$ Inertial
Modes in Various Power-law Density Models
\label{table}}
\begin{ruledtabular}
\begin{tabular}{ccccccc}
    & 
\multicolumn{2}{c}{$\beta=0$\footnotemark[1]}
& 
\multicolumn{2}{c}{$\beta=0.5$} 
&
\multicolumn{2}{c}{$\beta=1.0$} \\
$j$ & $\omega/\Omega_s$\footnotemark[2] & ${\bar Q}$
\footnotemark[3] 
& $\omega/\Omega_s$ & ${\bar Q}$
& $\omega/\Omega_s$ & ${\bar Q}$\\
\hline
4 & -1.2319 & 0.084 & -1.1607 & 0.0027 & -1.1224 & 0.013\\
  &  0.2319 & 0.031 & 0.3830  & 0.0031 & 0.4860  & 0.019\\
\hline
6 & -1.6434 & 0.018 & -1.5820 & 0.0023 & -1.5415 & 0.0022\\
  & -0.8842 & 0.018 & -0.8726 & 0.0077 & -0.8671 & 0.0059\\
  & 0.1018  & 0.006 & 0.1780  & 0.0036 & 0.2364  & 0.0031\\
  & 1.0926  & 0.017 & 1.1814  & 0.021  & 1.2408  & 0.018\\
\end{tabular}
\end{ruledtabular}
\footnotetext[1]{The stellar density profile is $\rho\propto 
(R^2-r^2)^\beta$.}
\footnotetext[2]{$\omega$ is the mode frequency in the rotating frame,
as in $\bxi\propto e^{im\phi-i\omega t}$.}
\footnotetext[3]{$Q_{\alpha,22}=\int\!d^3x\,\delta\rho_\alpha^\ast
\,(r^2Y_{22})={\bar Q}\hat\Omega_s^2$, with 
$\hat\Omega_s=\Omega_s (R^3/GM)^{1/2}$. $Q_{\alpha,22}$ is calculated 
with normalization $\int\! d^3x\,\rho\,\bxi\cdot\bxi=1$ and $M=R=1$.
For the uniform-density model, ${\bar Q}$ is calculated fixing
$\Gamma_1=1$ (otherwise the result is zero), while for other models we
use $\Gamma_1=d\ln P/d\ln\rho$.}
\end{table}

For NS models with nonuniform density profile, no analytical solution
for the inertial mode eigenfunction is generally possible.
Wu \cite{Wu05a} showed that if the star has a power-law density
profile, $\rho\propto (R^2-r^2)^\beta$, 
equation (\ref{eq:modeeqn}) is separable in the ellipsoidal
coordinates and one can easily obtain eigenfunctions that are accurate
to ${\cal O}(\Omega_s)$. However, the Eulerian density perturbation
caused by inertial modes remains small [$\delta \rho_\alpha/\rho =
(\omega_\alpha^2/c_s^2) \psi \sim \omega_\alpha^2/(GM/R^3) \nabla^2 \psi \ll 
\nabla\cdot {\bxi_\alpha}$], we expect that
$Q_{\alpha,22}$ is of order $\Omega_s^2$, similar to that for the
uniform density model ($\beta =0$).  Table I gives the mode
eigenfrequencies and tidal coupling coefficients calculated using the
eigenfunctions from \cite{Wu05a} and setting $\Gamma_1 = d\ln P/d\ln
\rho$. Corrections of order unity may arise when one is able to obtain
eigenfunctions accurate to ${\cal O}(\Omega_s^2)$.

\subsection{Odd-Parity Modes}
\label{subsec:odd}

For an inclined orbit ($\Theta\neq 0$), the $m=1$, odd-parity 
($j=2,4,\cdots$) modes can also be excited by the $l=m'=2$ tidal
potential (see \S IV). The relevant tidal coupling coefficient is
\be 
Q=Q_{\alpha,21}=\int\! d^3x\,\delta\rho_\alpha^\ast\,r^2Y_{21},
\ee
with $Y_{21}=-(15/8\pi)^{1/2}\sin\theta\cos\theta\, e^{i\phi}$.

The $m=1,~j=2$ mode is a r-mode. For a uniform, incompressible
star, the mode frequency is given by\cite{Saio82,Kokkotas99},
to order ${\cal O}(\Omega_s^3)$,
\be
\omega_\alpha/\Omega_s=-1-3\hat\Omega_s^2/4,\quad
\sigma_\alpha/\Omega_s=-3\hat\Omega_s^2/4.
\ee
The tidal coupling strength is\cite{holai99}
\be 
Q_{\alpha,21}=\left({3\over 8\pi}\right)^{1/2}{\hat\Omega_s}^2.
\ee

For higher order ($m=1,~j=4,6,\cdots$) modes, eigenfunctions including
$\Omega_s^2$ corrections are not available, and our results for
$Q_{\alpha,21}$ are estimates using $\delta\rho_\alpha=
(\omega_\alpha^2\rho^2/\Gamma_1 P)\psi$, with a $\psi$ that is
accurate to ${\cal O}(\Omega_s)$. The mode frequencies and 
coupling coefficients are listed
in Table II for the $j=4,6$ modes.

\begin{table}
\caption{Same as Table \ref{table} but for 
$m=1$ inertial modes
\label{table2}}
\begin{ruledtabular}
\begin{tabular}{ccccccc}
    & 
\multicolumn{2}{c}{$\beta=0$}
& 
\multicolumn{2}{c}{$\beta=0.5$} 
&
\multicolumn{2}{c}{$\beta=1.0$} \\
$j$ & $\omega/\Omega_s$ & ${\bar Q}$\footnotemark[1] 
& $\omega/\Omega_s$ & ${\bar Q}$
& $\omega/\Omega_s$ & ${\bar Q}$\\
\hline
4 & -1.7080 & 0.023 &-1.6620  & 0.001 & -1.6328& 0.005 \\
  & -0.6120 & 0.019 &-0.6511  & 0.009 & -0.6751& 0.032\\
  &  0.8200 & 0.140 & 0.9242  & 0.012 & 0.9897 & 0.052 \\
\hline
6 & -1.8617 & 0.005 & -1.8308 & 0.001 & -1.8088 & 0.001 \\
  & -1.3061 & 0.005 & -1.2941 & 0.006 & -1.2866 & 0.005\\
  & -0.4404 & 0.030 & -0.4833 & 0.017  & -0.4126 & 0.001 \\
  & 0.5373  & 0.075 & 0.6175  & 0.022 & 0.6726  & 0.019 \\
  & 1.4042  & 0.008 & 1.4431  & 0.016 & 1.4688  & 0.011 
\end{tabular}
\end{ruledtabular}
\footnotetext[1]{$Q_{\alpha,21}=\int\!d^3x\,\delta\rho_\alpha^\ast
\,(r^2Y_{21})={\bar Q}\hat\Omega_s^2$.}
\end{table}

\section{Effect of Inertial Mode Resonance}

The result of \S III (see Table I) shows that the coupling coefficient 
of an inertial mode to the $l=m=2$ tidal potential has 
the form $Q_{\alpha,22}={\bar Q}{\hat\Omega}_s^2$, with
${\bar Q}\lo 0.1$, and $\hat\Omega_s=\Omega_s(R^3/GM)^{1/2}$.
For most modes, ${\bar Q}$ is likely to be significantly smaller.
Given that our result for ${\bar Q}$ should only be considered
as an order-of-magnitude estimate (since we did not include
$\Omega_s^2$ correction to the mode eigenfunction), in the following 
we will scale our equations using ${\bar Q}\sim 0.1$. Similar consideration
applies to the tidal coupling of odd-parity modes (Table II).

An $m=2$ inertial mode is resonantly excited by the $m'=2$ tide when 
$\sigma_\alpha=\omega_\alpha+2\Omega_s=2\Omega_{\rm orb}$. 
We only need to keep the $l=2$ tidal potential,
with $W_{22}=(3\pi/10)^{1/2}$, $|{\cal D}_{22}^{(2)}|
=\cos^4(\Theta/2)$. Thus the GW phase shift due to the resonance is
\ba
&& \Delta\Phi_{\rm GW}=-2.55 {R_{10}^5\over M_{1.4}^5 q(1+q)}
\left({\Omega_s^2\over\varepsilon_\alpha\sigma_\alpha}\right)
\left({{\bar Q}\over 0.1}\right)^2\,{\hat\Omega_s}^2\nonumber\\
&&\qquad\qquad\times \left(\cos{\Theta\over 2}\right)^8~{\rm radian}
\label{eq:delphi}\ea
where $M_{1.4}=M/(1.4M_\odot)$, $R_{10}=R/(10~{\rm km})$.
Note that $\Omega_s/(\varepsilon_\alpha\omega_\alpha)\sim 1$, 
and ${\hat\Omega_s}=(\nu_s/2170~{\rm Hz})(R_{10}^3/M_{1.4})^{1/2}$
(where $\nu_s$ is the spin frequency). Thus, for $\nu_s\lo 300$~Hz,
$\Delta\Phi_{\rm GW}\lo 0.05$~radian (for $R\sim 10$~km) ---
such a phase shift is too small to affect GW detection.

Compared to the $m=2,~j=3$ r-mode (often called $l=m=2$ r-mode)
resonance studied in Ref.~\cite{holai99},
the phase shift associated with the inertial mode resonance 
(Eq.~\ref{eq:delphi}) is larger by a factor of 
$(a/R)^2\propto \Omega_s^{4/3}$, because the r-mode can only be excited
by the $m=2,l=3$ (Newtonian) tidal potential.

For an inclined orbit ($\Theta\neq 0$), the $l=m'=2$ tidal potential
$\propto Y_{22}(\theta_L,\phi_L)e^{-i2\Phi}$ has a component proportional
to $Y_{21}(\theta,\phi)$, and thus can excite $m=1$, odd-parity modes
at the resonance $|\sigma_\alpha|=2\Omega_{\rm orb}$.
For $\sigma_\alpha>0$ (i.e., the mode is prograde with respect to the
spin in the inertial frame), from Eq.~(\ref{eq:orbchange}) with $m=1$ 
and $m'=2$, and ${\cal D}^{(2)}_{12}=2\cos^3(\Theta/2)\sin(\Theta/2)$, 
we find that the GW phase shift associated with the resonance is 
\ba
&&\Delta\Phi_{\rm GW}=-10.21\,{R_{10}^5\over M_{1.4}^5 q(1+q)}
\left({\Omega_s^2\over\varepsilon_\alpha\sigma_\alpha}\right)
\left({{\bar Q}\over 0.1}\right)^2 {\hat\Omega_s}^2\nonumber\\
&&\qquad \times\left(\cos{\Theta\over 2}\right)^6
\left(\sin{\Theta\over 2}\right)^2{\rm radian}.
\label{eq:delp}\ea

For modes with $|\sigma_\alpha|\sim\Omega_s$ (such as the $m=1,~j=4$ 
modes), Eq.~(\ref{eq:delp}) has the same behavior as 
Eq.~(\ref{eq:delphi}) for the excitation of even-parity modes.
The exception is the $m=1,~j=2$ r-mode, which has $\sigma_\alpha/\Omega_s
=-(3/4)\hat\Omega_s^2$.
Since this mode is retrograde with respect to spin in the inertial
frame, we use Eq.~(\ref{eq:orbchange2}) to find
\ba
&& \Delta\Phi_{\rm GW}=162.5{R_{10}^5\over M_{1.4}^5 q(1+q)}
\left({\Omega_s\over\varepsilon_\alpha}\right)
\nonumber\\
&&\times \left(\sin{\Theta\over 2}\right)^6
\left(\cos{\Theta\over 2}\right)^2{\rm radian}.
\label{eq:delm1}
\ea 
Thus the phase shift is always very significant for $\Theta$ not
too close to $0$. However, the resonant condition $2\Omega_{\rm
orb}=|\sigma_\alpha|$ implies that the GW frequency at the resonance is
\be
\nu_{\rm GW}={3\over 4}{\hat\Omega_s}^2\nu_s=10.2\,{R_{10}^3\over M_{1.4}}
\left({\nu_s\over 400~{\rm Hz}}\right)^3\,{\rm Hz}.
\label{eq:nugw}\ee
Thus, only when $\nu_s>400M_{1.4}^{1/3}/R_{10}$~Hz can the resonance
occur in the LIGO band.

\section{Discussion}

The result in this paper indicates that for neutron stars
with spin frequencies $\nu_s\lo 100$~Hz, resonant
excitations of inertial modes during binary inspiral 
produce negligible phase shift $\Delta\Phi_{\rm GW}$ in the gravitational
waveform. At such spin frequencies, only the g-mode
resonance \cite{holai99} and the gravitomagnetic
resonance of the $m=2$ r-mode\cite{Flanagan06} may 
produce $\Delta\Phi_{\rm GW}$ of order 0.1~radian 
--- such a phase shift probably does not affect the 
matched filtering method for detecting gravitational waves.
Of course, one should keep in mind that the above number
is for ``canonical'' neutron stars with $M=1.4M_\odot$, $R=10$~km.
The phase shift depends strongly on the neutron star parameters:
For the g-mode resonance, $\Delta\Phi_{\rm GW}\propto R^{3.5}M^{-4.5}
\nu_{\rm GW}^{-1}$ (where $\nu_{\rm GW}$ is the gravitational
wave frequency at resonance, and it depends nonlinearly on the spin
frequency); for the gravitomagnetic r-mode resonance, $\Delta\Phi_{\rm
GW}\propto R^4M^{-10/3}\nu_{\rm GW}^{2/3}$ (where $\nu_{\rm GW}$
of the same order as the spin frequency). Thus the phase shift is larger 
for neutron stars with larger radii.

For higher spin frequencies ($\nu_s\go {\rm a~few}\times 100$~Hz), 
various resonances becomes important, including the f-mode
resonances studied in \cite{holai99} and the inertial mode
resonances studied in this paper. While most of these
resonances occur at the high-frequency band of the gravitational
waves ($\nu_{\rm GW}$ is of order the spin frequency, and thus
$\go {\rm a~few}\times 100$~Hz), it is of interest
to note that the excitation of the $m=1$ r-mode 
(the $m=1,~j=2$ mode in our convention; see \S III.B) occurs at 
the low-frequency band (tens of Hertz) of 
LIGO (see Eq.~[\ref{eq:nugw}]) and produces
very large phase shift (see Eq.~[\ref{eq:delm1}]).
While neutron stars with $\nu_s\go {\rm a~few}\times 100$~Hz
are found within binaries with a low-mass companion (e.g., white dwarfs),
it is not clear that they are produced in compact binaries
with another neutron star or black hole companion. If such
systems do exist, then the detection the $m=1$ r-mode resonance 
should be straightforward, and would lead to a clean constraint 
on neutron star parameters (radius and spin).

Finally, it is worth mentioning some caveats of the present study.
Our calculations of the tidal coupling coefficients for most inertial
modes are approximate, since we did not take into account of the 
${\cal O}(\Omega_s^2)$ correction to the mode eigenfunction.
More importantly, we have neglected buoyancy in the stellar models.
The Brunt-V\"as\"al\"a frequency inside a cold neutron star could
be as large as $100$~Hz \cite{Reisenegger92,Lai94}. 
While the finite buoyancy has a negligible
effect on the r-mode, it could affect the other inertial modes
in an appreciable way. It is worthwhile to study the property
(particularly the tidal coupling strength) of the inertial modes 
in nonisentropic neutron star models (see \cite{Yoshida00} and 
references therein).

\begin{acknowledgments}
This work was supported in part by NSF grant AST 0307252 (DL) and
the Natural Sciences \& Engineering Council of Canada (YW).
\end{acknowledgments}



\begin{thebibliography}{10}
\bibitem{comment}
In Ref.\cite{holai99}, mode decomposition was carried out in the
configuration space (i.e., $\xi=\sum_\alpha c_\alpha\xi_\alpha$),
which is in general not valid. Nevertheless, if one {\it assumes}
that the modes are orthogonal with each other, one obtains the same
expression for the energy transfer. See also \cite{lai97}
for a discussion of this issue.
\end{thebibliography}
\end{document}